\documentclass[12pt,preprint]{aastex}
\def\gtrsim{\mathrel{\hbox{\rlap{\hbox{\lower4pt\hbox{$\sim$}}}\hbox{$>$}}}}
\let\ga=\gtrsim
\def\lesssim{\mathrel{\hbox{\rlap{\hbox{\lower4pt\hbox{$\sim$}}}\hbox{$<$}}}}

 \def\deg=^{\circ}

 \slugcomment{Accepted to the Astrophysical Journal Sup. Series}
\begin{document}

                                \title{
Constraints on Intracluster Gas Bulk Motions in Clusters of Galaxies with {\sl ASCA}
                                }

                                \author{
                                Renato A. Dupke \& Joel N. Bregman
                                }

                                 \affil{
Department of  Astronomy, University of Michigan, Ann Arbor, MI 48109-1090
                                }
%===============================================================================
                                \begin{abstract}

The recent detection of Doppler shifts of X-ray spectral lines in the intracluster gas
of the Centaurus and Perseus clusters of galaxies with {\sl ASCA} has raised the question 
of the frequency of such velocity gradients and their relation to other 
cluster properties. To address this and improve the sample of ICM velocity gradient 
candidates we carried out a systematic search of the 
{\sl ASCA} archive for clusters with suitable observing conditions that allow 
full azimuthal mapping of gas velocities near the central regions.
Here we present the results of the azimuthal velocity distribution of a sample consisting of the 
``best'' observed clusters for velocity analysis.
Our sample encompasses
clusters with different physical and morphological characteristics, including both 
well behaved ``cooling flow'' and dynamically active clusters. The gas temperatures 
of the clusters in our sample range from 1.4 keV to 6.3 keV. 
We find velocity substructures in $\sim$ 15\% of the clusters in our sample on the order of 
a few $\times 10^3$~km~s$^{-1}$. The significance of the velocity 
gradients' is greater than 99\% in Abell 576 and RXJ0419.6+0225, the clusters where velocity gradients are 
most significant. The velocity gradients 
are consistent with transitory and/or rotational
bulk motion. The velocity gradients in these clusters are unlikely to be due to projected temperature 
anisotropies, {\sl ASCA} PSF, intrachip gain variations, and flux contamination from other sources. 
{\sl Chandra} images of these clusters  do not 
show obvious cluster alignments within the field of view covered by {\sl ASCA}. 
We also noticed a high frequency ($\sim$50\%) of anisotropic temperature gradients in the 
core of ``cooling flow'' clusters, suggesting the presence of frequent dynamical 
activity in the core of galaxy clusters.

                                \end{abstract}


%===============================================================================
                                \section{
Introduction
                                }

The merging of collapsed subsystems is believed to be the primary process of formation
(and growth) of galaxy clusters. Off-center merging imparts angular 
momentum to the intracluster gas, which can last several Gyr according to recent 
numerical and hydrodynamical simulations 
(Takizawa 2000, Gomez et al. 2002, Ricker et al. 1998, Motl et al. 2004).
The resulting kinetic pressure associated to this residual angular momentum in the intracluster
 medium (ICM) affects our
determination of physical quantities such as total cluster mass and baryon mass 
fractions, which are widely used to constrain cosmological parameters (e.g.
Bahcall et al. 1999; Allen et al. 2002). Intracluster gas bulk motions may also provide
clues to explain the rich phenomenology near the clusters core such as X-ray plumes 
(e.g. Sanders \& Fabian 2002),
radio halos (Carilli \& Taylor 2002)  and isophotal twisting (Mohr, Fabricant, \& Geller 1993). 
Recent {\sl Chandra} measurements of cluster 
gas temperature and surface brightness maps suggest the presence of bulk motions 
of $\sim$1000--2000~km s$^{-1}$ 
in the central regions of some galaxy clusters, e.g., Machacek et al. 2002 (Abell 2218), 
Markevitch et al. 2000 (Abell 2142), 
Vikhlinin et al. 2001 (Abell 3667), Kempner et al. 2002 (Abell 85). In some cases the velocities
inferred are as high as 4500 km~s$^{-1}$ (Markevitch et al. 2004). This suggests that 
bulk motions in the ICM may be common, arising from the processes of continuous 
cluster formation. Nevertheless, intracluster
gas bulk motions have been poorly studied until very recently, mostly due to the
technical difficulties involved in measuring ICM gas velocities directly. 
 
The most direct way to detect gas bulk velocities is
through the measurement of the Doppler shifts of X-ray spectral lines and, in particular, of the 
well defined FeK$\alpha$ line complex at $\sim$6.7 keV.
However, direct detection of intracluster gas velocities had to wait for the development 
of spectrometers with high spectral resolution and good instrumental gain stability, so 
that velocity changes of at least a few thousand km s$^{-1}$ could be detected reliably. 
Recently, Dupke \& Bregman (2001a,b)   
detected directly, for the first time, ICM bulk motions in the 
outer regions of the Perseus (Abell 426) and Centaurus (Abell 3526) clusters using {\sl ASCA} 
GIS data. They also measured velocity gradients, with greater significance, in the inner regions of 
the Centaurus cluster using the {\sl ASCA} SISs, which has been recently confirmed by {\sl Chandra} 
(Dupke \& Bregman 2005). The 
intracluster gas velocity distributions 
detected in these two clusters are consistent with systematic gas bulk rotation 
with correspondent circular velocities $>$ 1000 km s$^{-1}$, implying that a significant 
fraction of the intracluster gas energy can be kinetic. In both clusters there are other
indications of relatively recent dynamical turmoil (e.g. Stein et al. 1997, Ettori et al. 1998).

The initial results on Perseus and Centaurus prompted a larger systematic search
for velocity gradients in other clusters where such analysis is feasible. 
The direct measurement of radial velocities in the intracluster gas through Doppler 
shift of X-ray spectral lines is
limited by the instrument's spectral resolution, number of photons in a line and, in practice, 
by the spatial and temporal stability of the conversion between pulse-height and photon energy (gain) across the 
detectors. Typically, gain fluctuations, if not taken into
account, produce artificial velocities of the same order or greater than the velocities we 
are trying to measure. In order to 
compensate for intrachip gain fluctuations, one would like to have
multiple consecutive exposures of the full region of interest taken 
at the same CCD position. This is observationally expensive and prior knowledge about the 
velocity structure can significantly reduce the number of follow-up observations necessary to 
create a reliable velocity map. The prior knowledge about velocity distributions
can be obtained through the analysis of archived observations if
the instrumental gain is relatively stable and its variations are well-known. 
This is the case for the {\sl ASCA} satellite. 

The Advanced Satellite for Cosmology and Astrophysics ({\sl ASCA}) (1993-2000) was unique 
as the first mission
to combine high spectral resolution to wide frequency coverage and arcminute angular resolution. Therefore,
it is the first mission where ICM velocity gradients could, in principle, be measured. 
The {\sl ASCA} Solid 
state Imaging Spectrometers (SIS) have 
spectral resolutions similar to that of {\sl Chandra} and {\sl XMM} non-grating spectrometers. 
Furthermore, the instrumental
gain variations in position and time have been extensively studied, and can be included in
the uncertainties reliably. 
In this paper we present the azimuthal distributions
of intracluster gas velocities (and also gas temperatures and metal abundances) in a systematic search 
of the {\sl ASCA} archive for the clusters best suited for velocity measurements.

%===============================================================================

                                \section{
Data Reduction, Sample Selection, \& Methodology
                                }

\subsection{
Data Reduction \& Spectral Fittings
                                }
                                
{\sl ASCA} was equipped with four X-ray telescopes. 
The X-ray telescopes consisted of a set of four aligned identical telescopes. 
Each telescope was associated to a position-sensitive X-ray detector, two Gas-Imaging Spectrometers 
denoted by GIS 2 and GIS 3, and two Solid-State Imaging Spectrometers, denoted by SIS 0 and SIS 1. 
The Solid State Spectrometer (SIS) consisted of four 11 mm square, 420 x 422 pixel CCD (charged coupled 
device). It had a 22$^{\prime}$ x 22$^{\prime}$
field of view (four 11$^{\prime}$ squared CCDs) and covered a broad energy range (0.4 - 10 keV). 
The SIS originally achieved a FWHM energy of about 150 eV at 7 keV. 
The SISs could work in three clocking modes: 1-CCD, 2-CCD and 4-CCD, so that 
data from 1, 2 or 4 chips were read out for each detector. The gain stability of the SISs in 1-CCD mode
was significantly superior to that of the GISs (Dotani et al. 1997) and also to that using 2-CCD and 4-CCD modes.
Since gain stability across the detectors 
is very important for velocity studies, we only used data from the two SISs in 1-CCD mode for 
clusters with 
long exposure times. We selected observations where the cluster's center was located near the CCD 
center so that 
we could carry out an complete azimuthal analysis around the cluster's center.
For all pointings analyzed in this work, we selected data taken with high and medium bit rates, 
with cosmic ray rigidity
values $\ge$ 6 GeV/c, with elevation angles from the bright Earth of $\ge20^{\circ}$, 
and from the Earth's limb of $\ge10^{\circ}$ (SIS) using bright mode (SIS). We also excluded 
times when the 
satellite was affected by the South Atlantic Anomaly. Flickering pixels were removed 
from the data. 
We estimated the background from blank sky files provided by the {\sl ASCA} 
Guest Observer Facility. We used {\tt XSPEC} v11.2
software (Arnaud 1996) to analyze the X-ray spectra. 

The general geometrical configuration used to extract spectra for all clusters was chosen in such a way as
 to allow us to look for velocity gradients covering the region around the 
X-ray center. A 2$^{\prime}$ radius circle and 
several ``PIE'' sectors with a radial extent from 2$^{\prime}$ to 6$^{\prime}$ (or the CCD
border) were used for all clusters. The initial number of sectors was eight (corresponding to 
a sector angle of 45$^{\circ}$).  In some clusters of our sample 
many extraction regions had fitting parameters poorly constrained.
After it was clear that this was happening due to poor photon statistics we increased 
the ``PIE'' sectorial angle to 90$^{\circ}$.  The minimum number of sectors in
our analysis was four. If the velocity was still unconstrained, the cluster was excluded 
from our sample.
We also carried out the same spectral analysis with the PIE sector phased out 
by half the sector angle. This was done to check for ``borders''
where velocity gradients could be smeared out by the original region selection. This helped
us to test for spurious velocity gradients by looking at the smoothness
of the velocity field.

The spectra were fitted using the {\tt MEKAL} 
thermal emission model, based on the emissivity calculations of Mewe \& Kaastra (Mewe,
Gronenschild \& van den Oord 1985; Mewe, Lemen \& van den Oord 1986; Kaastra  1992),
with Fe L calculations by Liedahl et al. (1995). Abundances are 
measured relative to the solar photospheric values (Anders \& Grevesse 1989), 
in which Fe/H=$4.68\times10^{-5}$ by number. Galactic photoelectric absorption was 
incorporated using the {\tt WABS} model (Morrison \& McCammon 1983); Since there has been a decrease 
in X-ray efficiency of the SISs at low energy ranges since 
1994\footnote{heasarc.gsfc.nasa.gov/docs/ASCA/watchout.html, also Hwang et al. 1999}
we restricted the useful energy ranges of the instruments to 0.8--9.0 keV in all 
spectral fittings. Within this energy range the Hydrogen column density
(N$_H$) is not well constrained. There is a known degeneracy between the best-fit N$_H$ and gas 
temperature parameters. Since changes in best-fit temperatures may bias the determination 
of line centroids, leaving N$_H$ free to vary could produce significant
variations in the best-fit velocities. To check for possible 
velocity variations induced by this 
effect, the spectra from all regions were initially fitted with the N$_H$ parameter set free to vary.
We then averaged the N$_H$ obtained in the spectral fittings of all regions in a cluster 
and used this value to obtain
a set of spectral fittings where N$_H$ was fixed for all regions.
The absolute value that we chose to use in the spectral fittings with fixed N$_H$ is not important 
for our work since we are looking for {\it relative} bulk velocities
within a cluster. Since there were no significant variations of the velocity profiles 
with N$_H$ free and N$_H$ fixed we only present here the results for which N$_H$ was fixed at
the average value.

Spectral fittings using models such as {\tt MEKAL} are not very sensitive to small redshift differences within 
the variation ranges of the spectral parameter typical for galaxy clusters. Therefore, 
the minimization routines often gets stuck around local $\chi^{2}$ minima instead of the true minimum. This, if 
not checked carefully, may lead to inaccurate velocity measurements. To compensate for this, we implemented the 
following routine. We initially performed
spectral fits while varying the redshift values within reasonable ranges (with the command STEPPAR 
in {\tt XSPEC}) to find the "true" minimum. Then we fixed the redshift at that value, refitted the spectra 
to find the best-fit values of temperature, metal abundance, column density and normalizations. We 
subsequently used these results as initial input values in a new spectral fitting with the redshift 
parameter free to vary. From that we found 
the new "true" $\chi^{2}$ minimum and estimated the errors. This process was applied 
recursively until the best-fit redshift no longer changed.

\subsection{
The Inclusion of Intrachip Gain Fluctuations 
                                 }
                                 
The level at which we can constrain redshifts is not just a function of photon counts,
but also depends on other fitting parameters such as metal abundances (higher metal abundances 
leads to stronger lines and smaller errors)
and how well gas temperatures are defined. The accuracy of velocity measurements
also depends strongly on the variation of gain across the CCD. It is known that the SIS 0 and SIS 1 CCDs have different absolute 
gains, and this difference is typically
higher than the intrachip gain variation across each CCD. 
For a more detailed description of the gain variations of the {\sl ASCA} spectrometers see 
Dupke \& Bregman (2001a,b) and references therein. The intrachip variation across the 
CCDs used in this work (S0C1 and S1C3) are also different and each has a different dependence on time  
(Dotani et al. 1997\footnote{http://heasarc.gsfc.nasa.gov/docs/ASCA/newsletters/sis\_calibration5.html},
 Rasmussen et al. 1994\footnote{http://heasarc.gsfc.nasa.gov/docs/{\sl ASCA}/newsletters/sis\_performance2.html}).
 However, the variation of gain for both chips for different spectral lines is well known
 (Dotani 2002\footnote{http://www.astro.lsa.umich.edu/~rdupke/dotani.ps}) and can be reliably used 
 to estimate the systematic errors involved in measuring gas velocity gradients (Dupke \& Bregman 2001b).

 The standard deviation of the line centroid with chip position 
 in each time period can be derived from the documented gain time dependence determined 
 using the FeK$\alpha$ line at several (up to four) positions within each CCD 
 (Dotani et al. 2002). From that we determined the average value of the line centroid and the typical fluctuation of the measured redshifts for each period 
 ($\sigma_{gain} (CCD, t)$).
 To interpolate the values of $\sigma_{gain} (CCD, t)$ to intermediary time values we used 
 a linear fit. The results of the fittings are shown in Figure 1. It can be seen from Figure 1 that 
 no single CCD had continually the lowest gain variation. 
 The values of $\sigma_{gain} (CCD, t)$ 
 were used to weight the average of the best-fit velocities and errors measured separately with SIS 0 \& 1.
 
 [Figure 1]
 
 The event files analyzed in this work were corrected with the tool ASCALIN V0.9t, using the standard gain 
 calibration file {\it sisph2pi\_110397.fits}. In the later years of the {\sl ASCA} mission the Charge Transfer Inefficiency (CTI) evolved non-linearly with time\footnote{http://heasarc.gsfc.nasa.gov/docs/{\sl ASCA}/cti\_nonlinear.html}, especially for SIS 1 and this affected the overall chip gain. Its impact on
 intrachip gain variations is still not clear. The effect becomes noticeable for dates later than 
 mid-1998 for SIS 1 and mid-1999 for SIS 0. Most of the clusters for which we can measure reliable
 velocities were observed prior to that date so that their overall gain is not expected to have been
 significantly affected by the non-linear CTI variation. In order to estimate whether this effect could
 significantly affect the clusters in our sample, we repeated the velocity analysis for Abell 3558 using 
 the current CTI file patch released by GSFC {\it sisph2pi\_290301.fits}. We chose Abell 3558 because it 
 was the cluster in our sample that was observed most recently (Jan 2000). We show the
 results in Figures 6a,b. Even though there was a global gain shift by about 10\%, there was no noticeable
 change in the relative azimuthal distribution using different CTI files. Since we are looking for
 velocity {\it gradients} within the spatial range of one CCD, we do not care for the absolute values 
 of velocities. Therefore, global gain shifts are irrelevant to our analysis.  
 Since this is the cluster most likely to be affected by non-linear CTI variation we believe that we 
 are justified in using the standard gain calibration file for all of the clusters in our sample.

                                 \subsection{
 {\sl ASCA} PSF
                                 }

When observing regions within the detector field of view, the flux in the 
outer regions of the detector is contaminated by photons coming from the central regions 
due to the extended {\sl ASCA} point spread function (PSF) (Takahashi et al. 1995). This effect is 
energy dependent and becomes 
more significant for hot clusters. If we treat the PSF effect on the outer regions as a secondary 
spectral ``component'', PSF scattering will broaden the line, increasing 
fitting errors
and making velocity measurements more unconstrained.  However, technically, the situation is more
complicated because of the weak influence that the redshift parameter has in the global spectral fittings.
As mentioned in the previous paragraphs, if the continuum is changed the best-fit line 
centroid may be changed as well.
It is clear that if the cluster does not have a bona fide 
velocity gradient, the PSF should not create a spurious one directly but might create one indirectly 
by changing the continuum, and therefore, the best-fit gas temperatures, in the outer parts. This contamination,
if strong, would create a correlation of temperatures and velocities following the characteristic 
``cross'' shaped pattern of {\sl ASCA} PSF and should be relatively easily visible. To be conservative, 
in this work we disregarded any velocity gradients accompanied by a strong temperature gradient. 

If the cluster has a true velocity gradient (either rotational or transient) the 
PSF scattering will tend to artificially erase it by making velocity measurements of the contaminated 
regions imprecise. Since the PSF is energy dependent this effect could ``mask'' the presence of 
velocity gradients in hot clusters and may partially explain the upper limit in gas temperatures 
of our sample.  Since intracluster gas bulk flows and higher temperatures are both expected in 
merging systems, a fraction of these clusters with bona-fide velocity gradients 
may be obscured from our {\sl ASCA} analysis due to PSF scattering.

                                 \subsection{
 Random Velocity Scatter and its Dependence on Temperature   
                                 }
 
To test for other systematic uncertainties related to the methodology used to measure velocities 
we analyzed the 
simulated cluster spectra with different temperatures. We simulated 1000 clusters per temperature group, 
each group centered on a different temperature range (1.5 keV, 5 keV and 9 keV). We used the tool FAKEIT 
in XSPEC with column densities, metal abundances, and redshifts set to 5$\times$10$^{20}$cm$^{-2}$, 0.30 
Solar photospheric and
0.05, respectively. After simulating the clusters' spectra we determined the best-fit redshift using 
the same 
procedure that was applied to the real clusters. We show the results in Figure 2. There, we plot the 
best-fit velocities for each temperature group. The scatter along the temperature axis is due to the initial 
input ranges
which were chosen to be slightly different for display purposes and should be neglected. In this exercise we are interested in 
the scatter of best-fit velocities and its variation, if any, with temperatures. We also used a high value 
nominal number counts (270 kcount) since we do not want the standard statistical errors to dominate the 
uncertainties. In Figure 2 we show the standard deviation of the best-fit redshift values found ($\sigma$).
$\sigma$ shows a dependence
on temperature being a factor of $\sim$ 1.8 higher for a  $\sim$ 1.8 increase in temperature towards higher 
values and also increasing by $\sim$ 20\% for a factor of $\sim$ 3.3 decrease in temperature. The 
latter can probably be attributed to the velocities being measured primarily at lower frequencies by 
spectral lines other than the
FeK$\alpha$ . The former is likely to be associated to the relative weakness
of the line flux with respect to the continuum. These errors reflect to the ability of the current 
software to determine line centroids accurately and were added in quadrature to the statistical and gain 
errors for the clusters in our sample.

                                        \section{
Relevant Characteristics of the Sample
}

The sample analyzed here is built by elimination, based on suitability for velocity 
measurements. Aside from geometrical configuration and CCD mode we also performed an 
initial selection on the relevant parameters 
for velocity measurements such as counts within the spectral lines (or overall counts) and metal 
abundance. 
After finishing the initial selection we had about 30 clusters where 
velocity measurements were possible. However,  
half of the initial sample was later discarded due to the unreliability in 
measuring reasonably constrained velocities using both spectrometers. 
The final sample is
displayed in Table 1, where we also list other relevant characteristics of the clusters. 
 One can see that there is a significant interplay between metal abundances and total 
 number of counts that
still satisfies the constraints imposed for velocity measurements. This is not surprising 
since the counts that matters most in finding the line centroids are those within the 
Fe lines. 

Columns 1 \& 2 in Table 1 show the cluster names and average redshift (determined optically). The average gas 
temperatures and metal abundances over all regions analyzed including the central region are shown in 
columns 3 and 4, respectively. The average N$_H$ in the line of sight as estimated  
from HEASARC W3nH\footnote{http://heasarc.gsfc.nasa.gov/cgi-bin/Tools/w3nh/w3nh.pl} is shown in column 5.
Columns 6 \& 7 show the count rate and the total number counts within the CCDs in SIS 0 and SIS 1.
Previous indications of the presence of cooling flows is indicated in column
8. We list the number of years from the launch of 
{\sl ASCA} to the year that the observation was 
performed in column 9. In column 10 we show the position angle of the direction of North. 

Given the uncertainties involved in measuring velocities we, conservatively, used three criteria
to screen out clusters that have velocity gradients from our sample:
Our first criterion was that both instruments showed consistent results 
in the regions of interest, after taking into account global gain shifts. 
We show here (Figures 4-15) 
 weighted average plots of the 
 results obtained with both detectors. 
 To compensate for the dependence of the intrachip gain fluctuations with time, we weighted the 
 corresponding velocities and errors by $\sigma_{gain}^{-1}(CCD, t)$.
 No weighting was done
 for the average gas temperature and abundance since their uncertainties are not observed 
 to be significantly affected by intrachip gain fluctuations at the level that is found in the SISs.
 
 It may be argued that, instead of using the above mentioned procedure to build an ``average''
 velocity profile, one could fit the spectra of both SIS 0 and SIS 1 simultaneously using 
 data groups in {\tt XSPEC}. However, this would require a general correction for the global CCDs 
 gain difference, for example with
 the GAIN command in {\tt XSPEC}. However, based on our previous experience, we noticed that the 
 confidence level limits can be a function of how good this correction is. Therefore, we opted to fit the 
 spectral data separately and then use the above mentioned methodology 
to display the joined results, as done in previous works 
 (Dupke \& Bregman 2001b). In most cases the ``average'' results were 
consistent with the individual ones and the few exceptions are discussed below.
 
 Our second criterion was to screen the results based on an ``angular'' velocity resolution 
 criterion, i.e., only regions 
 (or a combination of adjacent regions) larger than 90$^\circ$ angular slices can 
 be used to consider the velocity gradient as non-spurious. This choice of a minimum angular slice
 corresponds roughly to the effectively angular resolution of {\sl ASCA} at the midpoint of the sector.
 The final criterion that we used to determine that the velocity gradient was robust was that 
 the velocity variations were not accompanied by significant changes of other spectral 
 parameters, mainly gas temperatures. This is due
 to avoid the best-fit velocities from being influenced by shifts in 
 the continuum, as described in section 2.3 and 2.4.

                                        \section{
Results: Best Candidates for Velocity Gradients
}
In Figures 3a-3l we show the smoothed 
SIS 0 images of the clusters in our sample to illustrate how the azimuthal angle relates to 
position angles. We plot our zero point, the direction to the true north and also the 
angular scale. All other angles mentioned henceforth will be given with respect to the zero points in
Figures 3a-3l increasing counterclockwise. When referring to a region we will cite 
the angle ``A'' at which the PIE sector is centered. The whole sector correspondent to 
that region would cover A-$\eta<$A$<$A+$\eta$, where $\eta$ is the minimum angular 
separation of the data points (either 30$^{\circ}$ or 45$^{\circ}$).

[FIGURES 3a-3l]

In Figures 4-15 we show the azimuthal distribution of the best-fit gas temperatures (TOP), metal 
abundances (MIDDLE) and redshifts (BOTTOM) for the clusters in our sample. The values plotted are
an average, which is linearly weighted by the inverse of the 1$\sigma$ gain variation 
($\sigma_{gain} (CCD, t)$) corresponding to each
detector. The errors of temperature and abundance represent 90\% confidence level ($\sim1.65\sigma$) 
and the velocity 
errors are $\sim$ 68\% confidence limits (1$\sigma$). The horizontal lines represent the corresponding 
confidence error limits
for the central circular region 2$^\prime$ in radius. 
In order to highlight the clusters with significant velocity gradients it is useful to plot the deviation
from the average, normalized by the errors (Figure 16).  
The clusters where velocity gradients are most robust 
are Abell 576 and RXJ0419.6+0225. We can see in Figure 16 that both
clusters have regions with weighted deviations equal or greater than 90\%. 
Below we discuss in more
detail the significance of the velocity gradients in these two clusters.

Abell 576 is a richness class 1 cluster with low central gas temperatures and average metal 
abundances. It has an optical redshift of 0.0389. The cluster shows mild indications of ``arm'' to the 
West of the cluster by 1$^{\prime}$--2$^{\prime}$
in the X-ray image (Figure 3a).  The temperature distribution of Abell 576 (Figure 4) shows marginal evidence of an 
anisotropic (elongated) ``cold core'' 
which is more pronounced towards the regions $\sim$ 60$^{\circ}$ and 270$^{\circ}$. In both regions
the temperature
drops from 4.25$\pm$0.35 keV to 3.55$\pm$0.15 keV in the center (90\% errors). The fractional 
errors in abundance measurements are larger and the overall distribution does not show significant 
gradients. The velocity profile, however, shows a significant velocity gradient. The redshifts of the 
regions encompassing $\sim$ 0$^{\circ}$ to 100$^{\circ}$ are significant higher than those found in the 
other regions. If we average (error weighted) the 
values of the four
regions with high redshifts (0$^{\circ}$, 30$^{\circ}$, 60$^{\circ}$, 90$^{\circ}$) and compare to the average 
value of the other regions 
(central region excluded), we obtain a velocity difference of 
(10.6$\pm$4.2)$\times$10$^3$~km~s$^{-1}$ (1$\sigma$ confidence limits).
The errors represent the statistical fitting 
errors and also include gain uncertainties ($\sigma_{gain} (CCD, t)$), 
 temperature--velocity scatter discussed in section 2.4 and also a correction for 
 the redshift scatter due to ability of a single
 temperature model to recover redshifts from possible multiple temperature components 
 discussed in section 4.2 below.

 [FIGURE 4 HERE]

In order to estimate the significance of the maximum velocity gradient we used the F-test. 
In applying the F-test we simultaneously fitted 
 spectra from SIS 0 \& 1 of the two regions with the highest best-fit discrepant velocities and compared $\chi^2$ 
 variations due to the change in the number of degrees of freedom. 
We compared the $\chi^2$ of fits which 
assumed the redshifts to be the same in the two projected spatial regions
to that of fits which allowed the redshifts in the two regions to vary independently, reducing 
the number of degrees of freedom by one. 
Since there are no significant changes in the best-fit redshifts obtained by different
instruments for the same region we locked the redshifts of SIS 0 \& 1 together 
within each of the two regions tested. 

For Abell 576 we chose the regions correspondent to 30$^{\circ}$ and 210$^{\circ}$ to apply the F-test.
The results show that these regions have different velocities at the 99.993\% confidence level. 
In Figure 17 we show the 
68\% (inner), 90\%(intermediary) and 99\% (outer) confidence contours for two interesting parameters 
(redshifts) of these two regions as well as the line correspondent to equal redshifts. In Figures 18a,b
we show the unfolded spectra (weighted by the photon energy) for both regions obtained with SIS 0 and SIS 1.

[FIGURES 17, 18a,b]

The high velocity gradient found here in Abell 576 adds to the body of indirect evidence to the presence
of high dynamical activity in this cluster. Rines et al. (2000) determined the mass profile of A576 
using the infall
pattern in velocity space for more than 1000 galaxies in a radius of 4 h$^{-1}$Mpc from the cluster's
center. They found that the mass of the central Mpc was more than twice of that found from
X-ray measurements, suggesting that non-thermal pressure support may
be biasing the X-ray derived mass. Their results are also in agreement with those of Mohr et al.
(1996).
Using galaxy photometric data, Mohr et al. (1996) found a high velocity tail (8 galaxies) separated by 
$\sim$ 3000 km/s from the cluster's mean. This high velocity population, however, is not clearly geometrically 
separated from the rest of the cluster. Mohr et al. (1996) suggested that they may be due to a past merger 
event in A576. Rines et al. (2000) more extended analysis found also a high galaxy concentration towards A576 
separated by $\sim$ 8000 km/s.
Kempner et al. (2003) analyzed the {\sl Chandra} observations of the core of this cluster and found 
sharp edges corresponding to jumps in gas density \& pressure roughly in the N-S direction, 
and suggest that the core substructures
are caused by a current merger during its second core passage.
 
RXJ0419.6+0225 is a nearby (z=0.0123) cool cluster discovered in the ROSAT 
All Sky Survey. It is among the brightest clusters in the 0.1-2.4 keV band, with the 
X-ray flux $\sim$ 4.5$\times$10$^{-11}$ erg~cm$^{-2}$s$^{-1}$ in the 0.1-2.4 keV band. It is 
dominated by a single elliptical galaxy NGC 1550. This cluster was observed by {\sl ASCA} 
for $\sim$80 ksec in 1999 in 1-CCD mode. We show the azimuthal distributions of gas temperature,
metal abundance and redshifts in Figure 5. The average temperature of this cluster is very
low ($\sim$1.38 keV), so that the redshift measurements is not significantly influenced by 
the FeK line complex (Figure 19a,b). 
The regions around 135$^{\circ}$ have
lower redshifts than the other regions. The velocity difference between the averages of low redshift 
regions (90$^{\circ}$, 135$^{\circ}$, 180$^{\circ}$) and the high redshift regions 
(0$^{\circ}$, 45$^{\circ}$, central) is
(4.2$\pm$1.5)$\times$10$^3$~km~s$^{-1}$  (1$\sigma$ confidence limits). The errors cited 
include the same corrections as those cited previously for A576. Since the radial distributions 
of temperature and abundances are flat throughout and we do not see evidence of a ``cooling flow'' 
we also included the best-fit values for the central region with those of the high redshift 
regions when estimating 
the velocity differences.
The notation is the same as that used in the previous paragraph and 
$\sigma_{gain} (CCD, t)$ used here is derived from the Si line. We used the Si line to estimate
the gain variations for this cluster because the intracluster gas temperature in this
cluster is very low and
the FeK line complex is barely visible (Figures 19a,b) so that the redshift is heavily biased towards the
FeL complex and other low energy lines, such as Si and S. As discussed in detail in Dupke \& Bregman 
(2001b), the gain fluctuation is energy dependent and velocity measurements with and without the FeK complex often 
give significantly different values. Therefore, we looked for the gain fluctuation correction that
corresponded to the lower energy range lines. The closest calibrated line to the FeL complex is that of 
Silicon at $\sim$1.8~keV.  

To perform the F-test we chose regions 0$^{\circ}$ and 180$^{\circ}$. The F-test shows that the velocities
of these two regions are different at the 99.36\% confidence level. In Figure 20 we show 
the confidence contour plots and the line of equal redshifts. 

[FIGURES 19a,b, 20]

In Figure 16 we see that Abell 376 also has regions with consistent high internal velocity deviations 
from the mean.
In Abell 376 there is a marginal ($\sim 3.2\sigma$) velocity gradient with 
higher than average velocities in the 90$^{\circ}$ to 
225$^{\circ}$ regions. Despite the apparent gradient, only two non consecutive regions; 135$^{\circ}$ and 
225$^{\circ}$ show significant variations (when compared to the 315$^{\circ}$ and central regions) (Figure 7). 
 We performed the F-test to estimate the significance of the discrepant 
regions 315$^{\circ}$ and 225$^{\circ}$. The confidence contours
are plotted in Figure 21 along with the line of equal redshifts. The difference is significant at 
less than 97.8\%. The 99\% confidence contours are degenerate and we do not have enough
statistics to confirm or refute the velocity gradient in this cluster with the available observation.
The temperature distribution in Abell 376 also shows indications of a mild azimuthal gradient, where 
the regions 30$^{\circ}$--90$^{\circ}$ have lower temperatures than the regions 
with azimuthal angles 
above 180$^{\circ}$ by 0.92$\pm$0.31 keV (Figure 7). Metal abundances are not very well 
constrained and are consistent with a constant value. 

[FIGURE 21]

                                        \subsection{AGN Contamination}
We are studying regions near the core of galaxy clusters where central dominant galaxies are usually 
found and are often active X-ray emitters. Therefore a question that may appear is if the X-ray emission
scattered from the central AGN can influence the redshift measurements through changes in the continuum. 
The contamination from the central AGN is analogous (and in general less important) than the contamination
from the intracluster gas in the central 2${^\prime}$ and was already discussed previously in 
section 2.3. 
However, bright background AGNs could still be a concern when analyzing cluster regions away from the center, 
where cluster emission drops significantly. To estimate the contamination level from background AGNs 
we analyzed archived 
{\sl Chandra} images of Abell 576 and RXJ0419.6+0225. Although there are no bright sources within the 
boundaries of the I-chips in RXJ0419.6+0225, we found a few bright sources in the field of view of Abell 576. 
The brightest AGN is located in one of the sectorial regions where we found the best-fit 
redshift to be significantly above the average.
However, {\sl Chandra} analysis showed that the AGN's flux accounts for less than 5\% 
of the total within the frequency range observed with {\sl ASCA} in a region similar (slightly smaller) to 
that used to extract spectra with {\sl ASCA}. 
Furthermore, its spectrum is typical 
of other AGNs and is well fit by an absorbed power law with index $\sim$2 and is very soft. 
AGNs do not influence 
the results presented in this paper, given their low fluxes compared to the typical 
brightness of the clusters at the radial
distances used in this work. 
 
                                        \subsection{Multiple Temperature Components}

Given the relatively high frequency of anisotropic temperature distribution found in
our sample (see next section) it may be reasonable to inquiry how the velocity gradients are
affected by projection effects.
If there is a projected component with different temperatures from that 
of the main cluster in the line of sight
towards one direction but not towards others there could be, in principle, an
spurious redshift difference due to shifts in the temperature recovered
from a 1-temperature spectral model fitting, affecting the best-fit recovered redshift.

In order to assess the effect that a possible asymmetrical second
component can have on the measured redshifts we performed a large
number of spectral simulations similar to those used in section 2.4. We added a second temperature 
component to the simulated spectrum and used the same procedure that we used in the real clusters
to recover the redshifts with a single temperature MEKAL model. The relative normalizations of the 
second component followed the ratio of counts towards different azimuthal regions of the real clusters.
This hypothetical secondary component would also add additional flux to that of the main cluster. 
In the simulations we used the characteristics of A576 and RXJ0419.6+0225 for the primary component.
These two clusters provide good temperature range coverage for the
other clusters in our sample. To be conservative we chose the largest possible flux 
difference among all azimuthal sectors analyzed in each cluster as a basis to determine 
the normalization of the second component in the simulations. 

We performed 50 simulations per temperature bin and measured the standard
deviation of the best-fit recovered temperature.  The results are shown in Figures 22a,b, 
where we plot the recovered temperature obtained with a single temperature model. 
The redshift errors shown are the standard deviation of
the best-fit values over the 50 spectral fittings.
We also show the best-fit recovered redshifts and the reduced chi-squared, using a 1-Temperature model. 
When analyzing Figures 22 one should keep in mind that this exercise only 
makes sense when looking for redshift 
changes within the limits observed in the real clusters.
Furthermore, the reduced chi-squared should be acceptable, i.e., the effects
of the secondary temperature components should not be obviously ruled out. 
For A576, it can be seen that the best-fit redshift for the
single temperature model is very stable with no evidence of variation in a
wide range of temperatures for the secondary component (Figure 22a). Only when the
secondary component reaches very low values ($<$ 2 keV)  is the effect
in the redshift noticeable. However, this happens for recovered
temperatures outside those observed in the real cluster ($<$3 keV). In addition, $\chi_{\nu}^{2}$ 
in the regions where the redshift ``jumps'' is significantly worse and a 1-Temperature
model does not fit the data well.
Despite the overall wider variation of the recovered
redshift with temperature for RXJ0419.6+0225 the results are similar to those of A576. The
recovered redshift is stable within the regions that the recovered
temperatures correspond to the limits observed in the real cluster (1.32-1.5 keV). 
The average (over both clusters) best-fit velocity scatter is 600 km~s$^{-1}$, which has been 
included in quadrature in the errors of the velocity differences.
We conclude that the effects of secondary spectral 
components on the velocities measured does not create spurious velocity gradients as large as those 
observed in the clusters with significant velocity gradients.

                                        \section{
Azimuthal Distributions of the Other Clusters in Our Sample
}

 [FIGURE 5 HERE]

{\em Abell 3558.}---In Figures 6a we show the azimuthal distributions for Abell 3558. The temperature 
profile is anisotropic
and there is a temperature gradient, which is strongest along the direction connecting
the regions $\sim$90$^{\circ}\pm$30$^{\circ}$ (with gas temperature 5.03$\pm$0.24 keV) and 
regions $\sim$300$^{\circ}\pm$30$^{\circ}$ (with gas 
temperatures 6.36$\pm$0.36 keV). Metal abundances are consistent with being constant throughout. 
The velocity distribution
suggests the presence of a gradient with the redshifts of the regions 150$^{\circ}$ to 240$^{\circ}$
being higher than that of the central region. However, the significance of the velocity gradient is 
marginal (average regions 150$^{\circ}$--240$^{\circ}$ have velocities is 2.2$\sigma$ above the average).
$\sigma_{gain} (SIS0, t)$ for the time of this observation was already too high, which weakens 
the reliability of the velocity measurements (see Figure 6b). We also show in Figure 6b 
the results using a different CTI file patch released 
by GSFC {\it sisph2pi\_290301.fits}. Aside from the overall redshift change downwards by $\sim$ 0.005, 
we do not see significant differences for any regions except 300$^{\circ}$. The redshift values 
for 
the 300$^{\circ}$ region with new CTI file are mostly due to SIS 1. Since the surrounding points do not 
show consistent behavior leading to
a region 
of low redshifts we believe that data point is spurious. 

[FIGURE 6a,b HERE]

{\em Abell 2589.}---From the distribution of velocities 
in Abell 2589 (Figure 8) we see that the regions near (45$^{\circ}$) seem to have a higher redshift
than that of regions 135$^{\circ}$--180$^{\circ}$ and central. Although,
tantalizing, the significance of this ``spike'' in the joint SIS 0\&1 velocity profile is low ($<$2$\sigma$).
The temperature and metal abundances profiles are consistent with being flat.

[FIGURE 8 HERE]
 
 {\em Abell 2052.}---The temperature distribution of Abell 2052 shows a 
 significant (at the 90\% confidence level) symmetric gradient between the regions 
 0$^{\circ}$--150$^{\circ}$ and 180$^{\circ}$--330$^{\circ}$. 
 (Figure 9). 
  Consequently, the ``cooling flow'' is steeper also at higher azimuthal
 angles ($>$180$^{\circ}$). There is an abundance gradient towards the low angle regions
 (0$^{\circ}$--90$^{\circ}$), where the abundance decreases from the central value of 
 0.64$\pm$0.07 solar to 0.45$\pm$0.1 solar. The velocity distribution is consistent with 
 no bulk flows.

[FIGURE 9 HERE]

{\em Abell 2657.}---Abell 2657 (Figure 10) 
also shows a mild, but significant at $>$ 90\% confidence ``cooling flow'' toward the 0$^{\circ}$ to 135$^{\circ}$ 
direction. Its abundance 
profile is consistent with a flat distribution.
We do not find evidence of velocity gradients and the distribution of velocities is consistent with
a flat profile within the range of detection of the instruments.

[FIGURE 10 HERE]

{\em Abell 1650.}---We show the azimuthal distributions for Abell 1650 in Figure 11 . We do not notice a
central temperature decline within
the region studied. There are indications of a mild, but significant, anisotropy in the 
temperature profile. The average 
temperature of the 30$^{\circ}$--60$^{\circ}$ regions is 4.88$\pm$0.32 keV. This is significantly 
($>$90\% confidence limits) lower than
the average value of the other regions, 5.78$\pm$0.44 keV. The maximum temperature gradient is 
as high as 1.4$\pm$0.4 keV (see below). The
velocity profile (center excluded) does not show any consistent indication of velocity structures.

[FIGURE 11 HERE]

{\em Abell 2244.}---SIS 1 observations of Abell 2244 do not provide many data points where velocities 
can be reasonably 
constrained. However, SIS 0 spectra shows much better constrained velocities consistent with a flat
distribution. The joint plot shows temperatures consistent with a constant value. 
The metal abundance profile is also flat (Figure 12). 

[FIGURE 12 HERE]

{\em Abell 3158.}--- There are no significant structures seen in the distribution of 
temperatures or metal abundances (Figure 13). The velocity profile is also marginally consistent 
with a flat distribution. 

[FIGURE 13 HERE]

{\em Abell 644.}---We do not detect significant temperature gradients with respect to the
central region of Abell 644 
(Figure 14), but there are marginal azimuthal variations of temperature.
Metal abundance are consistent with a flat profile. 
The velocity distribution shows indications of low velocity regions near the 120$^{\circ}$--240$^{\circ}$.
However, these regions are seen only with the SIS 1 and are intermittent, so that there are no
two adjacent 
regions that show a significant difference from the average and the velocity fluctuations are likely 
due to local SIS 1 temporal gain fluctuations. Due to these inconsistencies we conservatively do not consider 
A644 as a strong candidate.

[FIGURE 14 HERE]

{\em MS1111.9-3754.}---The plots for MS1111.9-3754 show a zone (90$^{\circ}$--135$^{\circ}$) that has the
lowest temperatures, highest abundances and lowest redshifts. Since they seem to be correlated 
 (Figure 15), based on our selection criteria, we conservatively consider the velocity gradient
 as spurious.
 
[FIGURE 15 HERE]
                           \subsection{Anisotropic Temperature Distribution}

Temperature and density inhomogeneities in the core of ``cooling flow'' clusters has been found
by {\sl Chandra} in many clusters previously thought to be well behaved, e.g., Centaurus (Sanders \& Fabian 2002), 
Perseus (Fabian et al. 2000), Abell 1795 (Fabian et al. 2001), Abell 496 (Dupke \& White 2003), 
Abell 2052 (Blanton et al. 2002) and others). These anisotropies are partly believed to be due to 
the frequent presence
of ``cold fronts'' (Markevitch et al. 2000). The interaction between 
the central engine of the cD and the surrounding ISM/ICM may produce other features such as cavities 
(McNamara et al. 2000) and X-ray arms that change the gas temperatures anisotropically. 

The azimuthal distribution of gas temperatures within the central 
6$^{\prime}$ of the clusters in our sample is anisotropic in half of the clusters in our sample.
The characteristic maximum temperature differences and their respective 90\% confidence errors are 
$\sim$0.92$\pm$0.31 keV (Abell 376), 
$\sim$1.43$\pm$0.40 keV (Abell 1650), 
$\sim$0.46$\pm$0.13 keV (Abell 2052), 
$\sim$1.34$\pm$0.26 keV (Abell 3558), 
$\sim$0.48$\pm$0.26 keV (Abell 576) 
, all show evidence for the
presence of cold cores. This suggests the presence of 
high dynamical activity in cluster cores in ``cooling flow'' clusters, probably associated to either 
nuclear cold fronts generated by accretion of sub-clumps (Markevitch et al. 2000) cD ``sloshing'' in the bottom of the 
gravitational potential (Dupke \& White 2003) or by the interaction between the AGN in the inner
regions of the cD galaxy and the surrounding ISM/ICM (McNamara, et al. 2000).

                                        \section{
Summary 
}

 The combination of X-ray derived gas density and temperature mapping has provided indirect 
evidence of the presence of gas bulk motions in clusters of galaxies. Previously, 
direct detection of bulk velocities 
had only been found only for the Centaurus cluster at small and large scales (Dupke \& Bregman 2001b) and 
the Perseus cluster, at large scales (Dupke \& Bregman 2001a) with the {\sl ASCA} satellite 
and possibly at 
very small scales with {\sl Chandra} (Sanders et al. 2004) and {\sl XMM} (Andersson \& Madejski 2004). 
Here we have measured 
the azimuthal velocity profile
for all the clusters in the {\sl ASCA} archive for which velocity mapping could be performed with 
useful precision.
The characteristics of the instruments on-board {\sl ASCA} together with the particularities of the observations in the
{\sl ASCA} archive limits our sensitivity to detect velocity differences of $\ga$ 2000 km~s$^{-1}$, so that
only clusters with very high internal velocities, as well as, projected cluster alignments/infall 
can be measured.

We plot the maximal 
internal velocity splittings of the sample in Figure 23 with the correspondent 90\% errors. 
The best candidates for velocity gradients in our sample are Abell 576 and RXJ0419.6+0225. 
They show significant velocity gradients 
independently with SIS 0 and SIS 1, with no biases due to correlations with temperature or
abundance gradients. After the gain uncertainties are taken into account the residual 
velocity differences are $\ge$ 3.5$\times$10$^3$km~s$^{-1}$ (A576) and 
$\ge$ 1.5$\times$10$^3$km~s$^{-1}$ (RXJ0419.6+0225)
at 90\% confidence level. 
The F-test shows that the velocity gradients for Abell 576 and RXJ0419.6+0225 
are significant at the 99.993\% and 99.36\%  confidence levels. 

Despite the natural bias of our sample towards the brightest clusters this works suggests
that strong bulk motions are present in a non-negligible fraction of nearby clusters.  
The presence of bulk motions in the intracluster gas has fundamental impact 
in the determination of physical quantities such as gas mass and baryon fraction and may help 
explain the X-ray -- Gravitational lens mass discrepancy in the central regions of specific clusters 
(e.g. Allen 1998; Machacek et al. 2002). It can, indirectly, induce the ``$\beta$'' discrepancy in 
clusters by providing additional kinetic support to the thermal component 
(e.g. Evrard 1990; Allen et al. 1992, cf. Bahcall \& Lubin 1994). It is expected
that the frequency of clusters with velocity gradients $>$$\Delta$V falls off as $\Delta$V$^{-4}$ 
(Pawl, Evrard \& Dupke 2005), so that
future similar analysis with {\sl Chandra} \& {\sl XMM} should increase the sample by more than
an order of magnitude, since they will allow us to reduce the 
errors associated with gain fluctuations and measure velocity 
gradients with magnitudes as low as $\ga$750 km/s.

\acknowledgments  We thank
K. Mukai for providing information about {\sl ASCA} SIS gain calibrations that 
was crucial to this work. We thank the referee for the helpful suggestions. 
We are thankful to Jimmy Irwin, Nestor Mirabal \& Ed Lloyd-Davies for helpful discussions.
We acknowledge support from NASA Grant NAG 5-3247. This research made use of 
the HEASARC {\sl ASCA} database and NED. 

%===============================================================================
                                
	\clearpage		      
%===============================================================================
                                %\begin{figure}
                                %\title{
Figure Captions
                                %}
%1
                                \figcaption{
Variation of the standard deviation of the SIS gain with time from {\sl ASCA} launch. 
Each value is calculated
from a distribution of gain in several (up to four) positions across the CCD, for the FeK line 
using Cas A as a reference source. We also plot
the linear fit used to interpolate the values of $\sigma_{gain}$
for different clusters. The solid lines represent the SIS 0 CCD 1 and the dotted lines SIS 1
 CCD 3.                                }
%2
                                \figcaption{
Variation of the standard deviation of the best-fit velocity with average gas temperature.
One thousand clusters were simulated per temperature bin using SIS0 responses and the 
spectra were subjected to the same analysis as the real data. We show the standard deviation 
of the measured velocities as 1$\sigma$.
The statistical errors ($\sigma_{stat}$ were typically found to be 200 km~s$^{-1}$, 250 km~s$^{-1}$, 350 km~s$^{-1}$ 
for clusters with temperatures 1.5 keV, 5.0 keV and 9.0 keV respectively. 
                               }

%3
                                \figcaption{
SIS 0 (chip 1) smoothed image of the clusters shown in this work. X-ray surface brightness contours are 
overlaid. We indicate the origin (0$^{\circ}$) of the azimuthal angle used to 
plot velocities, temperature and metal abundances in Figures 4--15. The azimuthal angle increases
counterclockwise. We also show the North-South direction and the angular scale.}

%4
                                \figcaption{
Azimuthal distribution of Temperature in keV (TOP), Metal Abundances in Solar photospheric (MIDDLE), and 
Redshift in units of 0.01 (BOTTOM) as 
a function of the position angle (degrees) obtained from the analysis of SIS 0 \& 1 data for
Abell 576. The direction to 0$^{\circ}$ is shown in Figure 3a.  
For all plots dark solid horizontal lines represent the confidence limits
for the central region with a radius of 2$^{\prime}$. Errors for temperature and metal abundance
plots are 90\% confidence limits and for velocity (redshift plot) are 68\% (1-$\sigma$). The velocity plot 
shows a weighted average of the results obtained with SIS0 \& SIS 1 separately, using the inverse of the 
1-$\sigma$ variation
of the gain at the time of the observation as weights. In the spectral fittings the Hydrogen column density 
was left initially free to vary for all regions and spectrometers and subsequently fixed at the final average 
value for all regions and spectrometers. The data points shown for temperatures, abundances and velocities 
are only those for which well-constrained values were obtained in both SIS 0 \& SIS 1.}

\clearpage
%5
                                \figcaption{
Same as Figure 3 for RXJ0419.6+0225.                       }
%6
                                \figcaption{
(a)Same as Figure 3 for Abell 3558. (b) Same as (a) but using the new CTI file patch sisph2pi\_290301.fits.
                          }
%7
                                \figcaption{
Same as Figure 3 for Abell 376.                  
                          }
%8
                                \figcaption{
Same as Figure 3 for Abell 2589.                  
                          }
%9
                                \figcaption{
Same as Figure 3 for Abell 2052.                  
                          }
\clearpage
%10
                                \figcaption{
Same as Figure 3 for Abell 2657.                  
                          }
%11

                                \figcaption{
Same as Figure 3 for Abell 1650.                  
                          }
%12
                                \figcaption{
Same as Figure 3 for Abell 2244.                  
                          }
%13
                                \figcaption{
Same as Figure 3 for Abell 3158.                  
                          }
%14
                                \figcaption{
Same as Figure 3 for Abell 644.                  
                          }
%15
                                \figcaption{
Same as Figure 3 for Abell MS1111.9-3754.                  
                          }
%16
\figcaption{
Error Weighted Distribution of Relative Velocities. We show all regions,
except for those that include the central 2$^{\prime}$, for all clusters in our 
sample . The 68\% (long dashed lines), 
90\% (short dashed lines) and 99\% (dotted lines) distribution
boundaries are also shown. The clusters with the highest significant velocity structures are 
indicated by different symbols: Abell 576 (open circles), RX J0419+0225 (open squares) and Abell 376 
(Diamonds).
                          }
 
 \clearpage
%17
\figcaption{
Confidence contour plot for the redshifts measured for regions 30$^{\circ}$  and 210$^{\circ}$ (using both 
SIS 0 \& 1 for Abell 576. The three contours 
correspond to 68\% (solid), 90\% (short-dash) and 99\% (long-dash) confidence levels. The line of equal 
redshifts 
is also indicated. The contours are found from simultaneous spectral fittings of four data groups 
(30$^{\circ}$ SIS 0\&1 and 210$^{\circ}$ SIS 0\&1), and the redshifts values for both instruments are locked 
together within the same region.         
}
%18
\figcaption{
(a)Unfolded (multiplied by energy) SIS 0 Spectra for regions 30$^{\circ}$ (black) and 210$^{\circ}$ (gray) of  
Abell 576. (b) Same as (a) for SIS 1.             
}
%19
\figcaption{
(a)Same as Figure 18a for regions 0$^{\circ}$ and 180$^{\circ}$ in RX J419.6+0225 (b) Same as (a) for SIS 1.             
}
%20
\figcaption{
Same as Figure 17 for regions 0$^{\circ}$ and 180$^{\circ}$ in RX J419.6+0225.                  
}
%21
\figcaption{
Same as Figure 17 for regions 225$^{\circ}$ and 315$^{\circ}$ in Abell 376.                  
}

%22
\figcaption{
Redshift biases from fitting a 1-Temperature model to a 2-Temperature simulated spectra with the characteristics 
of Abell 576. The X-axis shows the 
simulated temperature of the second component. The main component has a temperature of 4 keV. The single
temperature best-fit values for redshifts, temperatures are shown in the TOP and MIDDLE plots. The reduced chi-squared
of these fits are shown in the Bottom plot. The errors indicated the standard deviation of the best-fit values over
50 simulations (per bin)(b) Same as (a) for RXJ0419.6+0225.                  
}
%23
\figcaption{
90\% confidence limits for the maximum velocity difference for each cluster analyzed in this work.                  
}
                                %\end{figure}

\clearpage
%\thispagestyle{empty}
 %=== Table 1 ====================================================================
\begin{deluxetable}{rrrrrrrrrr}
\rotate
\small
\tablewidth{-10pt}
\tablecaption{Clusters Characteristics}
\tablehead{
\colhead{Name} &
\colhead{Redshift}  &
\colhead{Temp.}  &
\colhead{Abund.} &
\colhead{Gal. N$_{H}$} &
\colhead{SIS0/1 Cnt} &
\colhead{Total SIS0/1} &
\colhead{Cooling} &
%\colhead{$\Delta$V} &
\colhead{Yr After} &
\colhead{PA\_PNT}  \\
\colhead{} &
\colhead{(Optical)}  &
\colhead{(keV)}  &
\colhead{(Solar)\tablenotemark{a}} &
\colhead{($10^{22}$cm$^{-2}$)} &
\colhead{Rate (cnt/s)} &
\colhead{kcnts} &
\colhead{Flow} &
%\colhead{} &
\colhead{Launch} &
\colhead{(degrees)}
}
\startdata
Abell 576     & 0.039 & 3.96  & 0.64 & 0.06 &0.63/0.50 & 28/22  & weak\tablenotemark{b,f} &3.1  &82.9   \\
RX J419.6+0225& 0.0123& 1.39 & 0.33 & 0.12 &0.64/0.49 & 40/30  & no\tablenotemark{c,e} &6.5  &281.8  \\
Abell 376     & 0.048 & 3.75 & 0.59 & 0.07 &0.36/0.29 & 12/9   &weak\tablenotemark{b}  &4.5  &294.0  \\
Abell 2589    & 0.041 & 3.60 & 0.59 & 0.04 &0.71/0.57 & 12/9   &no\tablenotemark{c}    &6.8  &110.0  \\
Abell 3558    & 0.048 & 5.57 & 0.40 & 0.04 &1.43/1.13 & 88/69  &weak\tablenotemark{b,d}&6.9  &253.3  \\
Abell 2052    & 0.035 & 3.11 & 0.55 & 0.03 &1.28/1.03 & 51/41  &yes\tablenotemark{d}   &3.9  &248.8  \\
Abell 2657    & 0.04  & 3.43  & 0.43 & 0.06 &0.66/0.53 & 29/23  &weak\tablenotemark{b} &2.8  &117.4  \\
Abell 1650    & 0.085 & 5.62 & 0.51 & 0.02 &0.91/0.74 & 48/39  &yes\tablenotemark{d}   &2.9  &261.2  \\
Abell 2244    & 0.097 & 5.09 & 0.37 & 0.02 &0.78/0.61 & 27/21  &yes\tablenotemark{d}   &5.5  &60.1   \\
Abell 3158    & 0.06  & 5.06 & 0.53 & 0.01 &1.00/0.80 & 33/27  &no\tablenotemark{c}    &3.8  &167.6  \\
Abell 644     & 0.07  & 6.32  & 0.36 & 0.07 &1.44/1.12 & 79/62  &yes\tablenotemark{d}  &2.2  &80.6   \\
MS1111.9-3754 & 0.129 & 5.17 & 0.46 & 0.09 &0.32/0.27 & 18/15  &                       &3.3  &74.3   \\

\enddata
\tablenotetext{a}{Photospheric}
\tablenotetext{b}{$<$50 M$_{\odot}$year$^{-1}$}
\tablenotetext{c}{Consistent with 0 M$_{\odot}$year$^{-1}$ }
\tablenotetext{d}{Peres et al. 1998}
\tablenotetext{e}{From this work}
\tablenotetext{f}{Kempner et. al. 2003}
\end{deluxetable}

                               \end{document}